\documentclass[superscriptaddress,prl,showpacs,preprint]{revtex4}

\usepackage{amsmath}
\usepackage{graphics}
\begin{document}

\title{JAMMING DURING THE DISCHARGE OF GRAINS FROM \\ A SILO DESCRIBED AS A PERCOLATING TRANSITION}

\author{Iker Zuriguel }
\affiliation{Departamento de F\'{\i}sica, Facultad de Ciencias,
Universidad de Navarra, E-31080 Pamplona, Spain.} 

\author{Luis A. \surname{Pugnaloni}}
\affiliation{ Procter Department of Food Science, 
University of Leeds, Leeds LS2 9JT, United Kingdom.}

\author{Angel Garcimart\'{\i}n}
\email[Author to whom correspondence should be addressed. E-mail: ]{angel@fisica.unav.es} 
\affiliation{Departamento de F\'{\i}sica, Facultad de Ciencias,
Universidad de Navarra, E-31080 Pamplona, Spain.} 

\author{Diego Maza}
\affiliation{Departamento de F\'{\i}sica, Facultad de Ciencias,
Universidad de Navarra, E-31080 Pamplona, Spain.} 


\vspace{1cm}
\begin{abstract}

\noindent
We have looked into an experiment that has been termed the ``canonical example'' of jamming: granular material, clogging the outlet of a container as it is discharged by gravity. We present quantitative data of such an experiment. The experimental control parameter is the ratio between the radius of the orifice and the radius of the beads. As this parameter is increased, the jamming probability decreases. However, in the range of parameters explored, no evidence of criticality
---in the sense of a jamming probability that becomes infinitely small for a finite radius--- has been found. We draw instead a comparison with a simple model that captures the main features of the phenomenon, namely, percolation in one dimension. The model gives indeed a phase transition, albeit a special one.
\\
\end{abstract}

\maketitle

\newpage
The suggestive idea that jamming is at the origin of a new state of matter, amenable to be described in a thermodynamic formalism, has been recently proposed \cite{Liu}. Our experiment is akin to the familiar experience of shaking a saltcellar; salt grains plug the outlet of the container due to the formation of arches. Shaking or tapping is needed to break up the arch and restore the flow after each jam. A similar situation is found in many industrial applications, such as dosage hoppers, and in various transport phenomena. Ad hoc laws have been put forward based on experience and observation \cite{drescher}, but the physics of arching is still poorly understood. A new vision of jamming has recently been put forward, relying on the concept of fragile matter, which encompasses not only granular materials but foams, unstructured glasses and other systems \cite{cates}. The common feature they share is that they support certain stresses, called compatible, developed as a response to the external forces, but are unstable against incompatible stresses. 

We have devised a simple arrangement to conduct a quantitative study of the jamming that halts a granular flow. Our assembly consists of a scaled cylindrical silo with a circular opening at the base. The bin is filled with monodisperse glass spheres (radius standard deviation is about 1\%). Beads of different sizes were used in separate trials, but always with a diameter larger than 1 mm in order to reduce the relative importance of disturbances -- such as humidity or electrostatic interactions -- against gravity. The diameter of the bin is larger than 30 bead diameters; in this way, the finite size of the system can be neglected, as reported in the literature \cite{hirshfeld} and checked specifically for our set-up \cite{size}. The silo is always filled with the same procedure (a distributed filling \cite{zhong}, pouring the grains rapidly in the silo) in order to obtain a repeatable compaction of the material (a volume fraction of about 0.59) from run to run. Besides, the pressure at the base of a silo is known to be almost constant as long as the height of the material above it exceeds a certain level \cite{duran}. Indeed, the results of our experiment are repeatable --irrespective of the amount of material in the silo-- provided that the height reached by the beads is larger than twice the diameter of the silo. We refill the silo from time to time so that this condition is met. We have used a transparent glass silo in order to check that a mass flow develops inside it \cite{zhong}; a funnel was therefore not formed during the experiments.

When the experiment begins, beads pour freely from the outlet due to gravity. Soon thereafter, however, if the diameter of the orifice does not exceed a few bead diameters, an arch is formed arresting the flow. An electronic balance beneath the silo measures the number of grains $s$ which fall between two successive jams. The size of the avalanche is stored in a computer and then the arch at the outlet is destroyed so as to trigger another avalanche. This is accomplished by means of a jet of compressed air aimed at the orifice from beneath the silo. In this way, the compaction of the granular material remains approximately constant during an experimental run -- shaking would instead pack the grains together more densely \cite{nowak}. 
Further information on the set-up will be provided elsewhere. We made sure that variables such as the pressure of compressed air or the material from which the bin is made have a negligible effect in the data that we are presenting \cite{pres}. The parameter that we control is the ratio between the diameter of the outlet orifice and the diameter of the beads, called $R$ in the following. Both the size of the orifice and the size of the beads may be changed; only the ratio $R$ is relevant.

For a fixed value of $R$, the data collected after a large number of avalanches -- several thousands typically -- are best viewed in the form of a histogram. In Fig. 1, $n_R(s)$, which is the number of avalanches consisting of $s$ grains for a given $R$, is presented for $R=3$ in a semi-logarithmic plot. 
From the histograms of all the orifice sizes studied ($1<R<4.5$) it is
possible to define the jamming probability $J(R,N)$ as the probability of
finding an avalanche of size less than or equal to $N$ grains for a given $R$.
That is to say, $J(R,N)$ is the probability that an orifice of relative size $R$ gets obstructed at least once before more
than $N$ grains fall through it. Then $J$ can be evaluated as

\begin{equation}
J(R,N) =\frac{\sum\limits_{s=0}^{N} n_R(s)} {\sum\limits_{s=0}^{\infty} n_R(s)}
\label{e1} 
\end{equation}

We have obtained $J$ for different values of $R$ from the experiment; the results are shown in Fig. 2. The curves in the graph show that the bigger the size of the orifice, the
smaller the probability that it gets blocked before an avalanche larger than $N$ grains falls. 
Also, $J$ grows with $N$ for fixed $R$. These data are consistent with those found in a two dimensional hopper \cite{to}, obtained with a fixed number of grains, much smaller than in our experiment.

One could conceivably speculate from Fig. 2 that $J$ would perhaps tend to a step function as $N\rightarrow\infty$. The question is whether for that limit the jamming probability is one below a certain size $R_c$, while it is zero above $R_c$. In other words, whether there is a finite value of $R$ beyond which the orifice would not jam at all for an infinitely large silo. We can assess this by evaluating the size of the orifice for which the jamming probability is one half --let us call it $R_0(N)$. This can be easily obtained from the fit of $J(R,N)$ (see Fig. 2). The particular fitting function for $J$ chosen to obtain $R_0(N)$ is almost irrelevant because the slope of $J$ is quite large at $R=R_0$. In Fig. 3a we plot $R_0$ versus $N$ in semi-logarithmic scale. It is unclear whether $R_0$ saturates. Several functions can be fit reasonably well to $R_0(N)$ (see Fig. 3a), some of which saturate for $N\rightarrow\infty$, while others do not. Values of $N$ much bigger than $10^5$ are experimentally inaccessible in our laboratory. Therefore we cannot conclude whether there is or not a critical value for $R_0$. The family of curves used to fit $J$ also depends on $\alpha$, which is related to the slope of $J$ at $R_0$. The inverse of $\alpha$ is the width of the transition region from $J=1$ to $J=0$. The values for $1/\alpha$ are plotted in Fig. 3b. It decreases with $N$ but it is not possible to ascertain if it cancels out for a finite value of $N$, as would be the case for a step function. Both parameters, $R_0$ and $\alpha$, are not related in a simple way. 

These results can be understood in the framework of a one-dimensional percolation model.
Let us call $p_R$ the probability that one grain gets past the outlet of size $R$ without blocking it. Assuming that these events are independent, an avalanche of size $s$ would consist of one jamming event, $s$ non-jamming events and another jamming event. The probability of this sequence is

\begin{equation}
n_R(s)=(1-p_R)^2 p_R^{s}
\label{e3}
\end{equation}
This also holds if one considers the grains falling in clusters, or groups, rather than one at a time. In this case, $p_R$ is the probability that a group of grains passes the outlet without jamming, \emph{i.e.} without forming an arch. If outgoing groups have $k$ grains  in average, the probability of finding an avalanche of $s$ grains would be
$n_R(s)=(1-p_R)^2 p_R^{s/k}$.
From Eq. \ref{e3} one obtains $log(n_R)=2 \cdot log(1-p_R)+ s \cdot log(p_R)$, so $log(n_R)$ has a linear dependence on $s$,
exactly as observed experimentally. The value of $p_R$ can be determined from the histogram. The exponential dependence of $n_R(s)$ on $s$ precludes the interpretation of these data in the framework of self-organized criticality (SOC) \cite{bak}, because there exists a characteristic avalanche size manifested in the exponential distribution of events. Conversely, experiments with rough grains in different experimental conditions \cite{frette} and numerical simulations in a two dimensional silo \cite{manna} yield a power law distribution -- a signature of SOC.

Incidentally, the exponential distribution is a compelling evidence that arch formation is an uncorrelated process. From the temporal point of view, this is also revealed by the autocorrelation function of the series of avalanches and by the first return map (see Fig. 1, inset), which show no sign of memory between consecutive avalanches.

From the avalanche size distribution given by Eq. \ref{e3}, the mean size avalanche $S$ can be explicitly calculated, yielding:

\begin{equation}
 S= \frac{ \sum\limits_{s=0}^{\infty} s^2 n_R(s) }{ \sum\limits_{s=0}^{\infty} s n_R(s) } =
\ \frac{1+p_R}{1-p_R} 
\label{e2}
\end{equation}

The mean avalanche sizes obtained from our data are shown along with the above expression in Fig. 4. 
Indeed the histogram given by Eq. (\ref{e3}) and the mean avalanche size (\ref{e2}) are those of a one-dimensional percolation model \cite{stauffer}, which leads to a second order phase transition, but only at $p_R=1$. This is marked by a divergence to infinity in the mean avalanche size $S$, which is a measure of the susceptibility of the system. 
Let us remark that the limit $p_R=1$ is unattainable, so such a phase transition would never be observed. 
In practice, however, $p_R$ could be made high enough (if the radius of the outlet orifice is big enough), so
that jamming would seldom be observed.


We can put forward the idea that the jamming transition bears an analogy with the glass transition. By increasing $R$ one selects larger bridges only, so $1/R$ could be thought of as the ``temperature''. In the glass transition, the characteristic viscous time is certainly infinite for zero temperature; for a finite temperature, it can be extremely large but a power law divergence is not found \cite{mezard}. In the same sense, the mean avalanche size (or the susceptibility, for that matter) is infinite for $1/R \rightarrow 0$, and could be very large for finite values, but there would not exist a critical exponent for $S$. 

In this paper, we have shown that jamming events during the discharge of a silo of glass beads are uncorrelated and show no evidence of self-organized criticality, in contrast with other avalanche studies \cite{frette,manna} which were performed in different geometries. Moreover, we showed  evidence that suggest that the classical claim of the existence of a critical outlet size beyond which no jamming occurs --\emph{i.e.} $J$ becomes a step function-- could be invalid 
in a thermodynamic sense.
Comparison between experiment and recent numerical simulations \cite{pugnaloni} could help to gather more understanding. Finally, other interesting aspects that have not been considered, such as the influence of the roughness of the beads and their packing fraction, also merit further investigation.

The Spanish Ministry of Science and Technology (Project BFM2002-00414), the Government of
Navarre and the University of Navarre have supported this work. L. A. P. acknowledges a fellowship from Fundaci\'on Antorchas (Argentina) and the British Council. We thank Dr. G. C. Barker for his helpful comments, and Dr. O. Dauchot for discussing his ideas with us.

\newpage
\vspace{-0.5cm}
\begin{small}

\end{small}

\pagebreak

\newpage
\begin{center}

\begin{figure}[th] 
\includegraphics{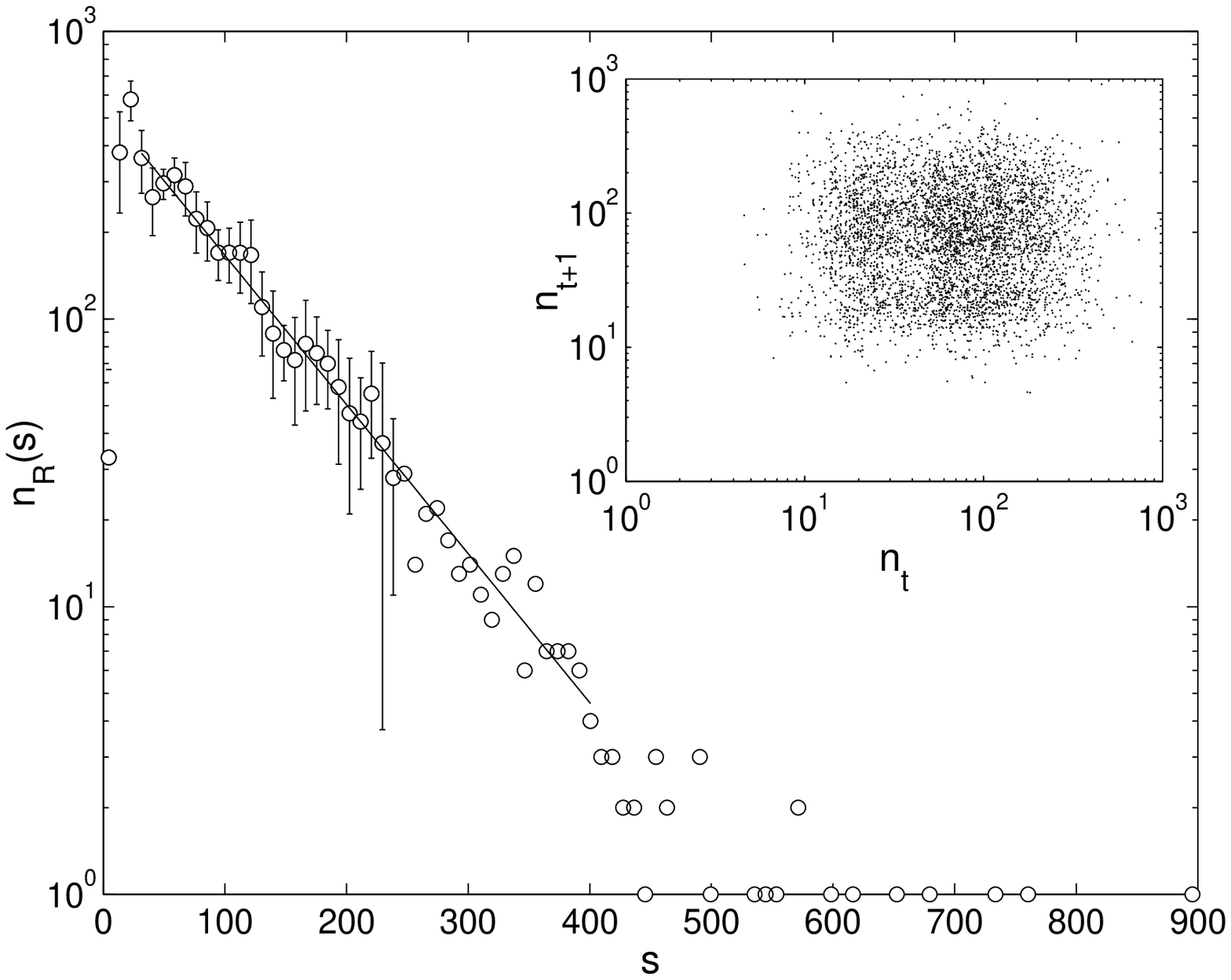}
\caption{\label{fig1} 
Histogram for the number of grains that flow between two successive jammings. Data correspond to $R=3$ (the beads have a diameter of 2 mm., and the circular orifice is 6 mm. wide). More than four thousand events have been recorded. The line is a linear fit. Inset: first return map, \emph{i. e.} the avalanche size $n_t$ versus the next avalanche size $n_{t+1}$. Note that $t$ is just a correlative index ordering the sequence of avalanches.
}
\end{figure}

\begin{figure}[th] 
\includegraphics{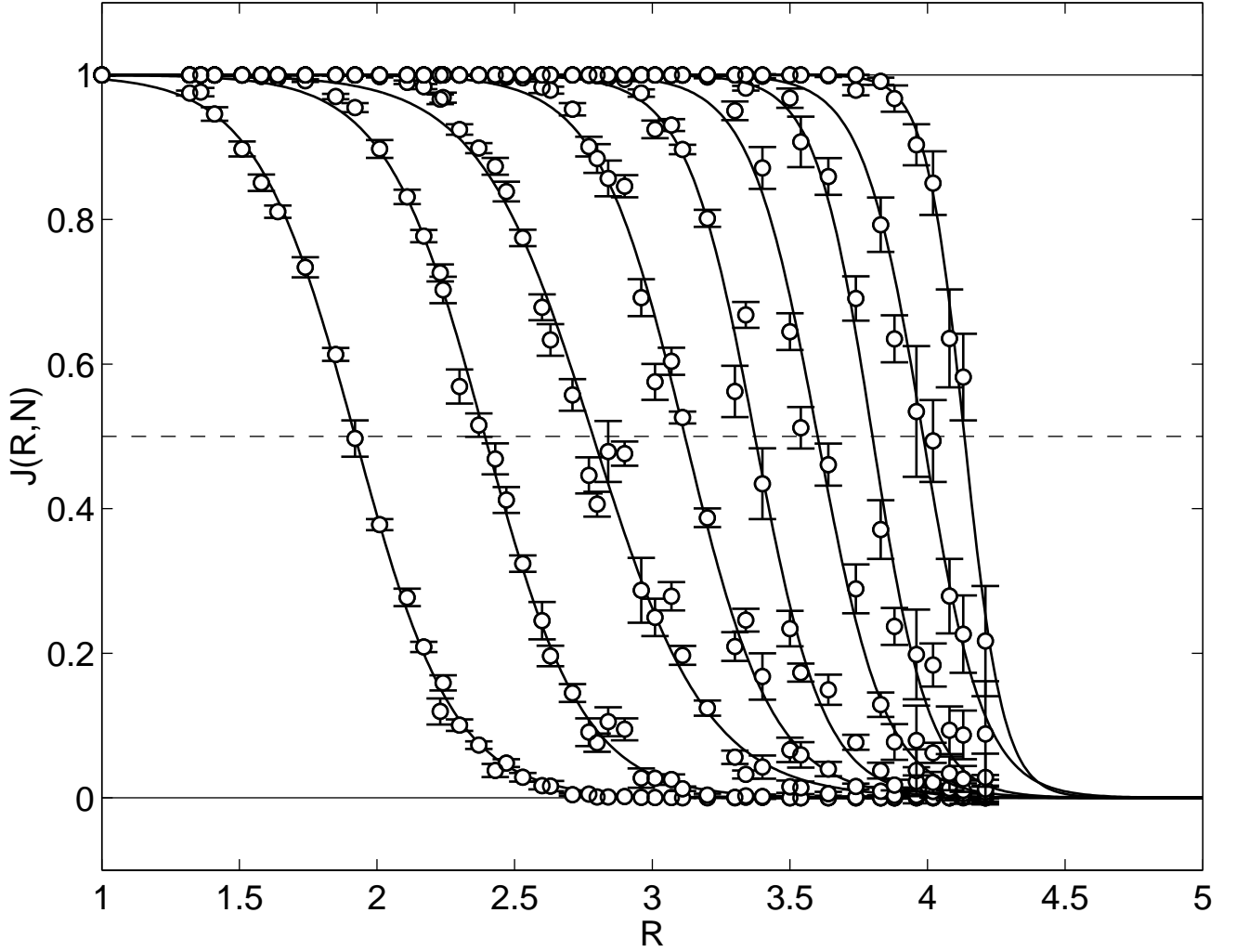}
\caption{\label{fig2}
Jamming probability $J$ as a function of $R$. Each series corresponds to a fixed $N$, i.e. the probability of the orifice getting jammed before $N$ grains pass through it. Data taken for several bead diameters are displayed. Succesive series, from left to right, correspond to $N=3, 10, 30, 100, 300, 1000, 3000, 10000 \text{ and } 30000$ beads. The fit corresponds to a hyperbolic tangent with two adjustable parameters: $J(R,N)=\frac{1-\text{tanh}[\alpha (R-R_0)]}{2}$ . The two parameters of this family of curves, $\alpha$ and $R_0$, are shown in Fig. 3. Error bars are statistical.
}
\end{figure}

\begin{figure}[th] 
\includegraphics{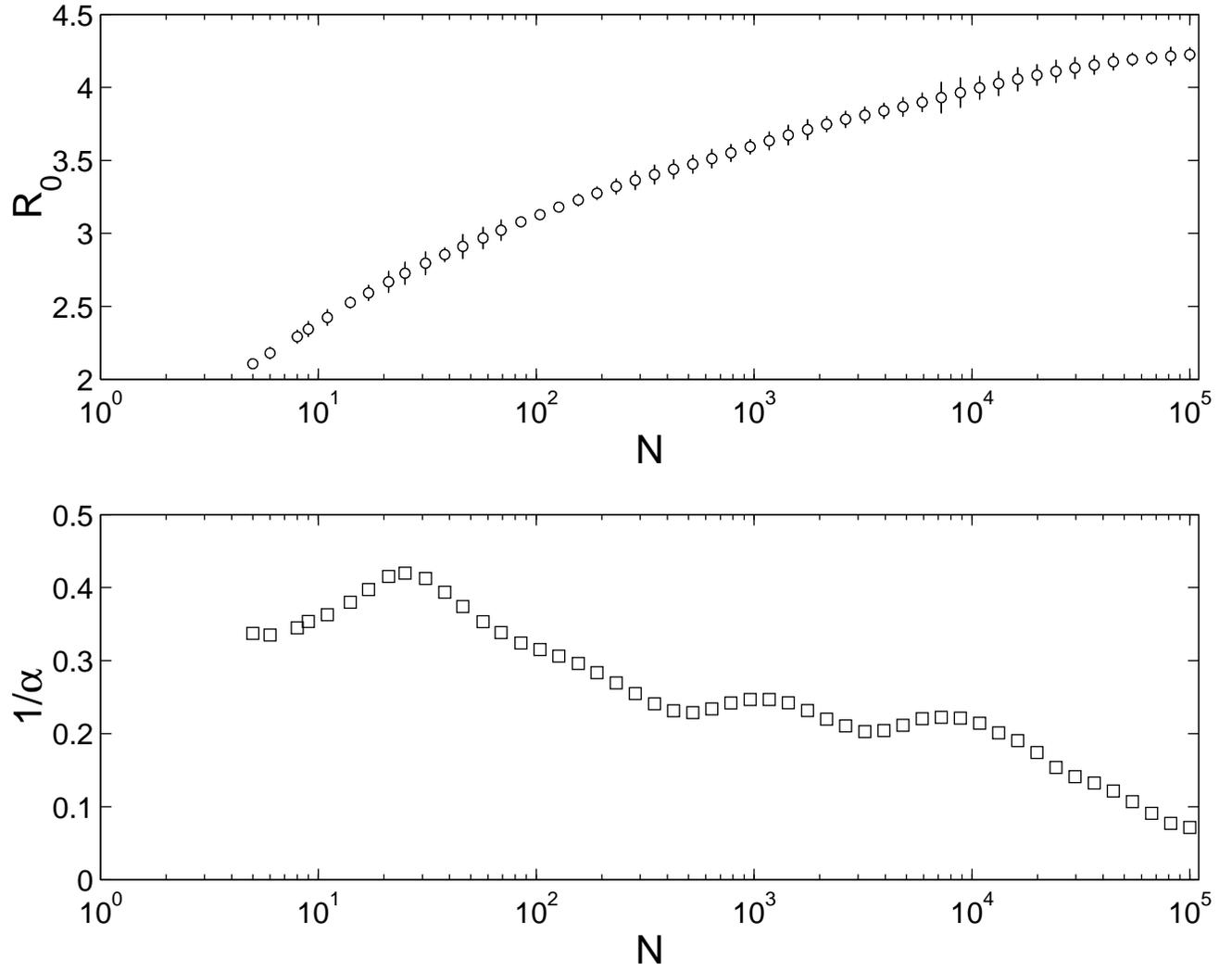}
\caption{\label{fig3}
(a) $R_0$ versus $N$. Note the logarithmic scale on the horizontal axis. (b) $1/\alpha$ as a function of $N$ in semilogarithmic scale. Both $R_0$ and $\alpha$ were obtained from the fits of Fig.2.
}
\end{figure}

\begin{figure}[h] 
\includegraphics{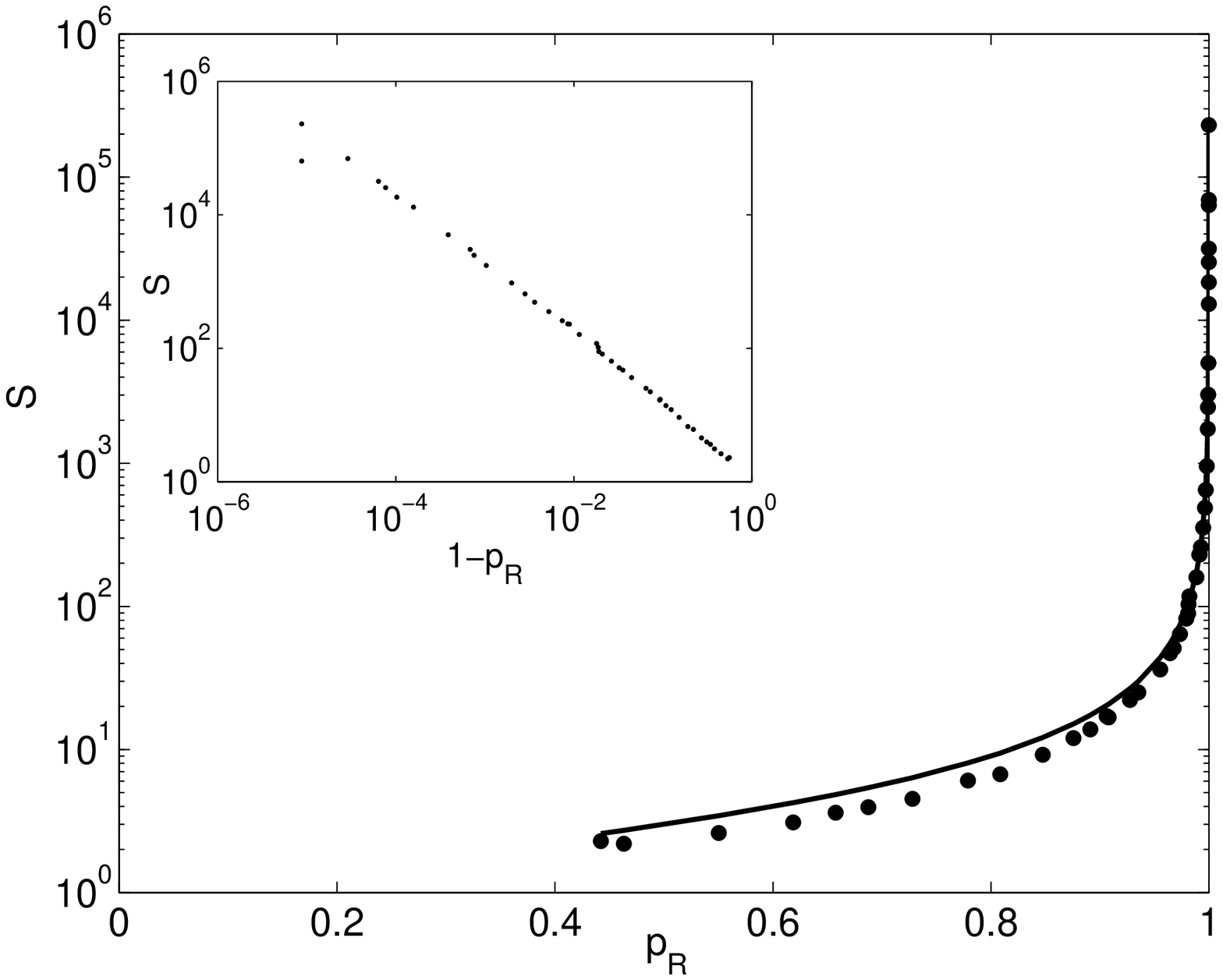}
\caption{\label{fig4} 
Mean avalanche size $S$ as a function of $p_R$: experimental data (points) and model (solid line). For the experimental data, $p_R$ has been estimated as $\frac{N_b}{(N_b+N_j)}$, where $N_b$ is the total number of fallen beads and $N_j$ the total number of jammings, for a given orifice size. Inset: the same data, plot in logarithmic scale as a function
of $1-p_R$.
}
\end{figure}

\end{center}

\begin{thebibliography}{}
\bibitem{size}
We have done this by performing experiments in several silos of different diameters, namely 30, 80, 120 and 150 mm, with beads of 2 and 3 mm in diameter. The results from the three larger silos are indistinguishable; not so for the smaller silo.
\bibitem{pres}
We set the pressure of the air jet to the minimum needed to break the arches, which is about 2 bar. We checked that
a pressure of up to 7 bar does not affect the results (the exponential tail of the histogram remains unchanged).
\end{thebibliography}
\end{document}